\documentclass[aps,amsfonts,prd,showpacs,nobibnotes,nofootinbib,%
tightenlines,twocolumn]{revtex4}
\usepackage{amsfonts}
\usepackage{epsfig}
\usepackage{latexsym}
\begin{document}
\hspace*{4cm}MIT-CTP-3397 

\title {QFT Limit of the Casimir Force}
\author{Marco Scandurra}\email{scandurr@lns.mit.edu} 
\affiliation{MIT Center for Theoretical Physics, 77 Massachusetts Avenue, Cambridge MA 02139}
\begin{abstract}
\noindent High precision  measurements of the Casimir effect and recent applications to micro electromechanical systems raise the question of how large the Casimir force can be made in an arbitrarily small device. Using a simple model for the metal boundary in which the metal is perfectly conducting at frequencies below plasma frequency $\omega_p$ and perfectly transparent above such frequency, I find that the Casimir force for plate separations $a<\lambda_p/2$, where $\lambda_p$ is the plasma wavelength is given by $\frac{-\hbar\omega_p^4}{24 \pi^2 c^3}$ which is independent of $a$. This result is considered the maximum value of the Casimir force for non-ideal metallic boundaries as calculated by quantum field theory. It differs from predictions of non retarded Van der Waals theory.  Implications of this result for geometries different from the planar one and in particular for the hollow metallic sphere are discussed.
\end{abstract}

\vspace*{-\bigskipamount}

\maketitle

\section{Introduction}
The Casimir effect, the attraction between parallel metallic plates in vacuum has undergone remarkable experimental verifications in recent years \cite{Bordag-Mohideen}. Perfect, grounded conductors in vacuum were predicted by Casimir in 1948 to attract with a force per unit surface given by
\begin{equation}
F\ =\  -\frac{\pi^2\hbar c}{240 a^4}, 
\end{equation}
where $a$ is the plates separation. Such formula, with opportune modifications,  was tested with increasing precision in experiments  featuring a sphere over a disk \cite{Lamoreaux},\cite{Mohideen} and parallel plates \cite{Carugno}. Experiments  aroused curiosity and  interest in this macroscopic manifestation of the quantum vacuum  among micro electromechanical and nano technology scientists who now seek practical applications  in miniaturized devices. It has been already shown that  Casimir forces can be used as mechanical actuators in MEMS \cite{Capasso}. Present technology seems to be able to overcome some of the historical problems related to Casimir experiments such as plates misalignment, and surface roughness. Ion beam smoothening, for instance, allows to reduce surface roughness to the atomic scale. It is expected that in the future the Casimir force will be tested at smaller plate separations. According to (1) two perfect conductors placed at a distance of 20 nanometers attract with a force of the order of ten thousands newton per square meter. Formula (1) indeed predicts an infinite force at vanishing plate separations, however in this regime the non-ideal behavior of conductors used in the laboratory becomes evident. Casimir's 1948 result is found by introducing a high frequency cut off in the Euler-Maclaurin summation formula \cite{Casimir}; after subtraction of the contribution of free unbounded space, what remains in the total energy is the Casimir term (1) and a series of terms depending on non negative powers of the cut off parameter $s$, such terms are neglected in the assumption that $s$  is small, i.e. that the conductivity is very large. In a plasma model for metals, the  requirement for the  validity of (1) is then  $a>>\frac{\pi c}{\omega_p}$, where $\omega_p$ is the plasma frequency of the metal.  It would seem that for smaller separations higher orders in the Euler-Maclaurin expansion could simply be attached, however they depend on the type of cut off function chosen and some ambiguity arises.  A detailed investigation of the interaction of the electromagnetic field with a smooth potential in the framework of renormalizable QFT would give a trustworthy result, such investigation is not simple: it is not clear how to model a smooth potential that  mimics a conductor. Corrections to Casimir's formula for non-ideal conductors were calculated by several authors \cite{Dzialoginski51,Hargreaves60}. In reference \cite{Bordag-Mohideen} the Casimir force for parallel plates is given with corrections up to the fourth order in terms of the plasma wavelength $\lambda_p=2\pi c/\omega_p$
\begin{eqnarray}
F  & = &  -\frac{\pi^2\hbar c}{240 a^4} 
\left[ 
1-\frac{16}{3}\frac{\lambda_p}{2\pi a} 
+ 24\left(\frac{\lambda_p}{2\pi a}\right)^2\right.\nonumber \\ 
   &   & -\frac{640}{7}   \left( 1-\frac{\pi^2}{210} \right)\left(\frac{\lambda_p}{2\pi a}\right)^3\nonumber\\  
   &   &   \left.+ \frac{2800}{9} \left( 1-\frac{163\pi^2}{7350} \right) \left(\frac{\lambda_p}{2\pi a}\right)^4 
\right] ;
\end{eqnarray}
this equation,  valid for separations between 100 $\mu$m and 100 nm,  assumes the penetration of the wave into the plates to be a constant equal to $\lambda_p/(2\pi)$. Formula (2) (adapted to the configuration of a sphere above a disk and corrected for finite temperatures\cite{Galina}) was tested in the laboratory at separations down to $\sim$ 70 nm.  At smaller separations physical chemistry predicts a static  interaction between atoms according to non-retarded Van der Waals theory \cite{London,Hamaker}. In this picture there is no scattering of the vacuum fluctuations  and  the speed of light is infinite; such theory predicts a force behaving as $-1/a^3$. It is of interest to investigate what quantum field theory has to say about this regime.  It is of interest to investigate   whether QFT  predicts   a finite limit for the Casimir force when boundaries fabricated with metals existing in nature are brought very close. In \cite{Bob2} a calculation of the scalar vacuum energy in the background of two gaussian potentials was calculated with the tools of renormalizable quantum field theory;  it was found that the Casimir energy does not diverge in the limit of vanishing separations, as long as the potentials are non singular. Lifshitz approached the problem of attraction between dielectric surfaces at small separations in 1956 \cite{Lifshitz}. A $a^{-3}$ divergence was predicted for vanishing separations and this result was proved to be substantially correct in a recent experiment \cite{Svedese}.  In the present article  I propose a simple way to model the metal boundary in the critical region $a=\lambda_p/2$ and below. I will assume a classical Drude model, and a simple hypothesis: the metal fully reflects waves below plasma frequency and fully permits higher frequency waves to pass through. Thus we can envision the electromagnetic vacuum as composed of two different sets of modes: 1) modes with frequency $\omega>\omega_p$ passing through the boundary without suffering any scattering or attenuation; these modes move unperturbed in space, they do not contribute to Casimir energies or Casimir stresses. 2) Modes with frequency $\omega<\omega_p$ fully reflected and trapped by boundaries, they  contribute to the Casimir effect whatever the topology of the boundary is.	This model allows to put the plasma frequency as the limit of integration and sum whenever a spectral summation of modes is required to calculate physical quantities.  In this model the metal plates are seen as passive scatterers and the Casimir force comes about through  the pressure exerted by the vacuum fluctuations; under this assumption a $a^{-3}$ divergence is untenable since very high frequencies are not stopped by the metal and should not contribute to the pressure.

\section{Wave propagation in metals}
Let us very rapidly review the classical electrodynamics of wave propagation inside metals. The wave vector in a conductor is given by
\begin{equation}
k^2=\frac{\omega^2}{c^2}\left( 1-i\frac{\sigma (\omega)}{\omega \epsilon_0}\right)
\end{equation}
where $\sigma(\omega)$ is the conductivity and $\epsilon_0$ is the permittivity of vacuum.  According to Drude's model the conductivity is given by
\begin{equation}
\sigma(\omega)\ =\ \frac{N_e e^2}{m_e}\left(\frac{1}{\nu+i\omega}\right)
\end{equation}
where $N_e$ is the density of free electrons, $m_e$ is the mass of the electron and $\nu$ is the frequency of collision with the lattice atoms, which for good conductors like silver is of the order of $10^{13}$ Hz. If the frequency $\omega$ of a wave entering a metal is lower than the collision frequency, the imaginary term in (4) can be neglected and  we are in the normal skin-effect regime, i.e. the fields possess a propagation term and a damping term governed by the length $\sqrt{2\epsilon_0 c^2/(\omega \sigma)}$. On the contrary, when $\omega>>\nu$, damping collisions with the atoms become insignificant and the electrons behave like a plasma. In most of Casimir experiments this is the relevant regime and we will deal with it for the remainder of this article. The wave vector becomes
\begin{equation}
k=\pm \frac{1}{c}\sqrt{\omega^2-\omega_p},
\end{equation}
where $\omega_p=N_e e^2/(m_e\epsilon_0)$. The electric field penetrating inside the metal along the direction $z$ is
\begin{equation}
E=E_0\left(e^{-\frac{z}{c}\sqrt{\omega_p^2-\omega^2}}\  e^{-i\omega t}\right)\ .
\end{equation}
We see that (6) contains no wave term as long as $\omega <\omega_p$. This means that a wave below plasma frequency is exponentially  damped inside the metal, but unlike the normal skin-effect  damping, there is no oscillatory term and the attenuation of the field is more rapid. When the $\omega$ approaches $\omega_p$ the metals becomes an ideal conductor and the field inside the metal vanishes. Finally when $\omega >\omega_p$ a wave term appears and the wave can propagate as in a dielectric. Given this kind of behavior one is led almost naturally to consider the metal as an ideal conductor for frequencies below plasma frequency and as an ideal transparent medium (n=1) for higher frequencies. And assumed that in traversing a perfectly transparent medium waves do not loose any energy or momentum, we can neglect them in the computation of Casimir forces and perform mode summation only up to the highest integer number corresponding to plasma frequency. This is of course a rough approximation, the advantages of the model, however  rest on the possibility of analytical calculations with qualitatively correct results.

\section{A high vacuum regime}
 Let us consider two parallel metallic plates immersed in the electromagnetic vacuum. The plates obey the model for conductivity described in the previous section, according to which the  modes with wavelength shorter than $\lambda_p$ do not see the boundaries at all and, being homogeneous and isotropic, do not contribute to the Casimir energy\footnote{In this picture it is neglected the energy associated with an isolated  boundary, see \cite{Bob}.}
The electromagnetic waves below plasma frequency obey conducting boundary conditions at the walls, thus inside the cavity only a discrete set of modes given by
\begin{equation}
k_{l,m,n}\ = \ \sqrt{\left(\frac{\pi l}{a}\right)^2+\left(\frac{\pi m}{a}\right)^2+\left(\frac{\pi n}{a}\right)^2}
\end{equation}
are possible, where $l,m,n$ are positive integer numbers  and the minimum allowed frequency inside the cavity is given by
$\omega_0=c\cdot\pi/a$. Clearly the only contribution to the Casimir effect is coming from those wavelength $\lambda$ obeying 
\begin{equation}
\lambda_p\ <\ \lambda\ <\ 2a\ .
\end{equation}
These waves are long enough to be reflected by the electron plasma of the metal and short enough to fit in the cavity as  cavity electrodynamics  requires.
If now we bring the plates closer, down to a distance $a<\lambda_p/2$,  no mode will be trapped inside the cavity anymore. While this happens inside in the outer region the long wavelength modes are still able to propagate with a continuum spectrum of frequencies, thus a sort of  high vacuum has been created inside the cavity. In this regime the Casimir pressure pushes the plates without encountering  resistance from the interior.
This ``high vacuum'' force  can be easily calculated:
\begin{eqnarray}
F_{h.v.} & = & - \frac{hc}{\pi^2} \int^{\omega_p}_0 dk_z\ \int^{\sqrt{\left(\frac{\omega_p}{c}\right)^2-k_z^2}}_0 dq\ q\ \frac{k_z^2}{\sqrt{q^2+k_z^2}}\nonumber\\
         & = & \frac{\hbar \omega_p ^4}{24\pi^2 c^3}, 
\end{eqnarray}
where $k_z$ is the component of the wave vector normal to the plates and a transformation to polar coordinates in the $x-y$ plane has been accomplished.
According to  (9) the force does not increase any further as the critical separation $\lambda_p/2$ is crossed.
Let us perform a more accurate calculation of the pressure in the case of a generic separation $a$, let us include the pressure in the interior $P_{in}$ and let $a$ go to zero after subtraction $P_{in}-P_{out}$. In the interior we have
\begin{equation}
P_{in}\ =\ \frac{hc}{\pi} \sum^{[\frac{a\omega_p}{\pi c}]} _{n=0} \int^{\sqrt{\left(\frac{\omega_p}{c}\right)^2-(\pi a/n)^2}} _0 dq\ q\ \frac{(\pi n/a)^2}{\sqrt{q^2+(\pi a/n)^2}},
\end{equation}
where the square bracket on the top of the summation represents the maximum integer smaller than $\frac{a\omega_p}{\pi c}$. The external pressure $P_{out}$
is given by eq.(9). After performing the summation in (10) the total pressure is found to be:

\begin{eqnarray}
P_{in}-P_{out} & = & \frac{ h \omega_p\beta}{12 a^3}\ + \frac{h\omega_p\pi \beta^2}{4 a^3}\ -\frac{h c\pi^2 \beta^3}{8 a^4 }\ +\ \frac{h \omega_p\pi \beta^3}{6 a^3}\nonumber\\ 
   &   & -\frac{h c\pi^2 \beta^3}{4 a^4}  -\frac{h c\pi^2 \beta^4}{8 a^4}                 
\end{eqnarray}
where $\beta=[\frac{a\omega_p}{\pi c}]$. This function is not continuous. If the energy is calculated a similar discontinuous expression is found; thus the force cannot be obtained as a derivative of the energy. Expression (11) is not suitable for analytic study as a function of $a$, however it has an analytic limit for $a\rightarrow 0$ and for $a\rightarrow \pi c/\omega_p$, in both cases: 
\begin{equation}
P_{a\rightarrow 0}\ =\ -\frac{\hbar \omega_p^4}{24\pi ^2 c^3}.
\end{equation}
which is the same value found in (9).

The r.h.s. of (12) is the maximum value that the Casimir pressure can take in an experiment with real conductors, it has a strong dependence on the plasma frequency of the metal, thus  slight changes in the number of free carriers of the metal are expected to produce significant changes in the force at small separations. For instance, copper and silver will show significantly different Casimir forces in this regime. The virtue of (12) is the lack of divergence at $a\rightarrow 0$.  In the table below the numerical value of the maximum Casimir force is given for several metals in Newton per square meter; the plasma frequency is shown in units of $10^{16}$ rad/sec.

\vspace{0.5cm}
\begin{tabular}{lcc}
       &\hspace{2cm}   $\omega_p$\hspace{2cm} & $P_{a\rightarrow 0}$ \\
       &               &      \\
Al     &  $1.75$   & $1500.29$ \\
Cu     &  $1.63$   & $1103.76$ \\
Au     &  $1.38$   & $576.99 $ \\
Ag     &  $1.36$   & $534.91 $ \\
\end{tabular}

\section{The spherical conducting shell}
The model investigated above  led us to a simple conclusion regarding the pressure at small plate separations: the pressure is inward. Such conclusion is indeed valid also for different boundary shapes and for closed cavity resonators. Among these the spherical shell is of particular interest since it is believed to possess a Casimir force directed away from the center\footnote{Recently this has been put into discussion \cite{Bob}} \cite{Boyer}. A hollow conducting sphere has, much like the parallel plates, a largest wavelength that is allowed in its interior, this is the first harmonic of the sphere. When the radius of the shell is made sufficiently small, the wavelength of the first harmonic is so small that the electron plasma is not able to stop the wave and confine it within the shell. Again, no contribution to the pressure can come from the inner space and this region can be ignored. Only low frequency modes in the outer space are scattered; then, if the pressure is to be ascribed to radiation momentum transfer it is directed toward the center of the sphere and the sign of the surface tension, unlike Boyer's 1968 result,  can be nothing but negative. That the effect of external mode scattering is a pressure directed toward the center, is suggested by experiments in which gold, silver and iron microparticls are repelled by laser light\cite{keiji}. Such pressure can be calculated by means of scattering cross sections in the Rayleigh regime. A pencil of radiation of intensity $I$ incident on a tiny sphere contributes with a radiation pressure
\begin{equation}
 P_k\sim -C (I/c) k^4 R^4,
\end{equation}
where $C$ is a numerical factor and the sign minus indicates that the pressure is directed toward the center. Integration of (13) over all possible states $k$ gives the total pressure for $R\rightarrow 0$. Such pressure  behaves as $-R^{4}$. In Boyer's paper \cite{Boyer} the energy is found to depend on $+R^{-1}$, which implies a pressure behaving like $+R^{-4}$. If Boyer's result and the conclusions of this article are both correct, a slope of the kind shown in Fig.1 should be the likely description of the pressure for small radii. Note the existence of a minimum and a vanishing force for values of the radius close to $\lambda_p$. Unlike the case of parallel plates the pressure does not reach a finite value for  $R\rightarrow 0$. It is important to say that this calculation considers the shell as infinitely thin. In an experiment the thickness of the shell is at least of the order of $\lambda_p$; when the radius becomes small this thickness is not negligible, therefore, strictly speaking, Boyer's result is not relevant in this situation.

\begin{figure}[ht]\unitlength1cm
\begin{picture}(6,6)
\put(-0.5,0){\epsfig{file=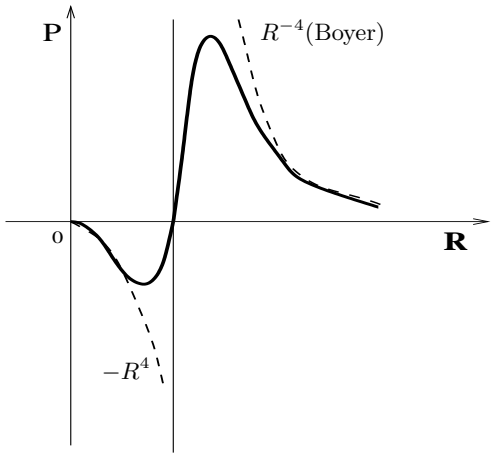,width=6.5cm,height=6cm}}
\put(2.9,5.5){$R^{-4}$(Boyer)}
\put(0.8,1.0){$-R^4 $}
\end{picture}
\caption{Casimir pressure on a metallic spherical shell with plasma frequency $\omega_p=2\pi/\lambda_p$. The upper dashed curve represents Boyer's pressure for a shell with infinite conductivity, the lower dashed line represents the pressure for vanishing values of $R$, when the conductivity is given by a step function} 
\end{figure}

\section{Conclusions}
In this article a simple way to model a conductor in vacuum  was presented.  An  upper limit to the modulus of the Casimir force was given, this limit is found imposing a sharp integration limit, in mode summation, given by the plasma frequency of the conductor. In a plot of the conductivity $\sigma(\omega)$ this would correspond to an infinite step function with foot at $\omega_p$. The Casimir force was calculated for parallel metallic plates with a separation smaller than the plasma wavelength. In this regime the modulus of the Casimir force takes the  value (12) which  does not depend on separation and does not diverge for $a\rightarrow 0$.  In the case of a spherical metallic shell, the surface tension is negative,  i.e. the sphere tends to collapse when the radius drops below a critical value; in this regime the surface tension increases with the radius, while in Boyer's calculation the tension is inversely proportional to the radius. Similar considerations can be made for cylindrical shells, cubes and parallelepipeds. In all these geometries the Casimir pressure in the high vacuum regime is necessarily negative and its modulus can be calculated by simply ignoring the interior of the cavity.
The results of this article are in disagreement with non-retarded Van der Waals theory which predicts and attraction of the type  $-1/a^3$. Such behavior is clearly absent in (12). The reason for this is probably in the assumption of a classical, static and non self interacting background. Such an assumption, which is typical of  Casimir-like calculations, does not describe correctly the physics of the system at small separations where the particle nature of the boundary becomes evident. 
One thing however should be stressed, the effect of force attenuation for a$\leq \lambda_p/2$ found in this work is independent of the fact that the molecules of the metal are very close: it is not an effect due to the overlapping of the boundaries, one could in fact use plates with a very small plasma frequency, such us semiconductor plates, and the effect of force attenuation would appear already at distances of the order of microns where Van der Waals forces play no role.  Another application of the model outside the regime of small separations is in a large cubic box of volume $V$, whose walls are made of a conductor with plasma frequency $\omega_p$; equation (9) predicts that the box traps a finite energy 
\begin{equation}
E_{box}\ = V\ \frac{\hbar \omega_p ^4}{8 \pi^2 c^3}\ ;
\end{equation}
here the walls of the box are at macroscopic distance from each other and non retarded forces play no role.

\vspace{1cm}
\noindent {\bf \Large Acknowledgments}\\

\noindent I wish to thank  Robert Jaffe for encouragement and helpful discussions. This work is supported in part by funds provided by the U.S. Department of Energy (D.O.E) under cooperative research agreement DF-FC02-94ER40818

\end{document}